\documentclass[12pt]{article}
\begin{document}

\begin{center}
{\Large \bf Some field-theoretical aspects of two types of the Poincar\`{e} group representations}

\vspace{0.5 cm}

\begin{small}
\renewcommand{\thefootnote}{*}
L.M.Slad\footnote{slad@theory.sinp.msu.ru} \\

\vspace{0.3 cm}

{\it Skobeltsyn Institute of Nuclear Physics,
Lomonosov Moscow State University, Moscow 119991, Russia}
\end{small}
\end{center}

\vspace{0.5 cm}

\begin{footnotesize}
The capabilities of some approaches to the relativistic description of hadronic states with any rest spin are analysed. The key feature in the Wigner's construction of irreducible representations of the Poincar\`{e} group which makes this construction fruitless in the particle physics is picked out. A realization of unitary irreducible representations of the Poincar\`{e} group of the standard type, which yet have never been considered, is discussed. The viability of the description of hadrons by the Poincar\`{e} group representations of the standard type in the space of the infinite-component ISFIR-class fields is pointed out.
\end{footnotesize}

\begin{small}

\begin{center}
{\large \bf 1. Introduction}
\end{center}

Introducting the notion of the electron spin as of its inherent angular momentum with which a magnetic moment is inseparably connected \cite{1}, 
\cite{2}, has served as a bright response to a number of laws established by that time by the experimental physics. In the relativistic classical theory, it was proposed to associate the spin with an antisymmetric tensor $s^{\mu\nu}$ \cite{3}, \cite{4}. In the nonrelativistic quantum theory, Pauli \cite{5} has put the electron spin in correspondence with the matrix operators connected with the three-dimensional rotation group. The first relativistic wave equation for the electron with spin, that have entailed triumphal consequences for the quantum theory of the electromagnetic and weak interactions of leptons, has been obtained by Dirac \cite{6} by splitting the Klein--Gordon equation with the results of \cite{5} taken into account.

The problem of relativistic description of particles with any spin yet has no solution which would to the full extent meet the present experimental picture of hadronic states \cite{7}. In this context, we wish to draw attention to two types of field-theoretical realization of the symmetry generated by the Poincar\`{e} group transformations which can be finally connected with two mathematically different types of the Poincar\`{e} group representations  possessing also essentially different potentials for such a description.

The realization of one such type belongs to Wigner \cite{8} who held the opinion that the Poincar\`{e} group representations can, to a large extent though not entirely, to replace the quantum mechanical equations. Relating the wave vector of this or that particle with spin to the space of an unitary irreducible representation of the Poincar\`{e} group, Wigner has formulated the classification of such representations and the prescript on inclusion of the spin in their structure on the basis of his "small group" concept. These classification and prescript are widely known, and, with some variations, are set forth in many subsequent works by various authors, from which we mark out an article by Bargmann and Wigner \cite{9} and the monographs \cite{10} and 
\cite{11}. However, the attempts to apply Wigner's formalism of Poincar\`{e} group representations to construct a field theory real particles of nonzero spin, or even to obtain the Dirac equation for the electron, are poorly presented in the literature if exist at all. This situation looks extraordinary enough in view of the well-established efficiency of various symmetry approaches in their physics applications. To find the reasons for the absence of the mechanism for obtaining the field equations, currents and Lagrangians at Wigner \cite{8} and at his followers is one of the aims of the present work. 

The course of searching for these reasons is partially given in parallel with preparations for realizing another, standard, type of the Poincar\`{e} group representations, for doing that we concentrate attention on a number of aspects of representations of subgroups of the Poincar\`{e} group. First of all, it is the question of the separate and joint realization of representations of the rotation group $SO(3)$ (and the proper Lorentz group $L^{\uparrow}_{+}$) in the space of functions of the coordinates (and of the time) ${\cal L}_{f}$ and in an abstract space ${\cal L}_{a}$. Then we touch equations possessing symmetry with respect to the Poincar\`{e} group transformations. In our assertions, we are guided by the elegant description of all irreducible representations of the proper Lorentz group and of all linear relativistic-invariant equations given by Gelfand and Yaglom \cite{12}. In this description, the spin associates with a matrix antisymmetric tensor operator, that represents a natural extension of the nonrelativistic Pauli analogue \cite{5} and conforms to the classical description proposed in \cite{3}, \cite{4}. The relativistic-invariant equations of Gelfand and Yaglom include, in particular, the equations of Maxwell, Dirac, Weyl (for two-component fields) \cite{13} and of Fierz and Pauli (for spin 3/2) \cite{14}. Unfortunately, the paper \cite{12} and the monographs \cite{15} and \cite{16} made on its basis have not received sufficient publicity. They have not been followed by comprehensive field-theoretical researches.

The realization of the Poincar\`{e} group representations in the spaces of solutions of the relativistic-invariant equations carried out in the present paper is accompanied by the proof of two conclusions: (1) there exist an infinite number of nonequivalent irreducible representations of the 
Poincar\`{e} group, all being characterized by the same values of the Casimir operators connected with the mass and rest spin of a particle, i.e. the indicated spin and mass values do not characterize an unique irreducible representation of the Poincar\`{e} group, but rather an infinite class of such representations; (2) the space of solutions of every relativistic-invariant equation divides into a direct sum of spaces of unitary irreducible representations of the Poincar\`{e} group. We do not aspire after completeness in the description of such a realization, because the relativistic-invariant equations and Lagrangians contain all the necessary information for particle physics. At the same time, it is essentially important that, aside from a very specific Wigner's type of irreducible representations of the Poincar\`{e} group which, being fruitless in the particle physics, plays in it the role of "the false sun" $,$ there exists the standard type including, in particular, all approved group-theoretical aspects of quantum mechanics and electrodynamics.

In the framework of the discussion of the Wigner's realization of irreducible representations of the Poincar\`{e} group, we first of all note that in the paper \cite{8} the notion of spin has no equitable description in all inertial reference frames, because its transformation under any element of the group 
$L^{\uparrow}_{+}$ reduces to a suitable transformation in the rest frame of a particle. Then we pay close attention to the key feature of the Wigner's construction which does not take place in the standard definitions of representations, though does not contradict them. It consists in the dependence of the transformation parameters associated with the transition from one inertial reference frame to another on the argument of the vector-valued function subjected to this transformation. This feature makes impossible the introduction of a relativistic spin operator and matrix operators which are necessary for constructing the linear relativistic-invariant equations, various currents and Lagrangians.

In conclusion, we present arguments in favour of the opinion that the reducible representations of the Poincar\`{e} group of the standard type realized in the space of solutions of the linear relativistic-invariant equations have good prospects to be suitable for the description of hadrons. These arguments are connected with the analysis, begun in the 2000s, of the theory of the ISFIR-class fields which transform under the $L^{\uparrow}_{+}$-group representations decomposable into an infinite direct sum of finite-dimensional irreducible representations. The existence of versions of such a theory of free fermionic fields with mass spectra consistent with the picture expected in the parton bag model of hadrons is proved in \cite{17}, \cite{18}, \cite{19}.

The chain of our reasonings on two types of the Poincar\`{e} group representations necessarily includes the assertions already known to a greater or lesser extent, however, we accompany them by accents and comments leading us to the declared goal. First of all it concerns the representations of the rotation group $SO(3)$, because the completeness and the perfection of their existing description are commonly recognized. At the same time the most meaningful sides of this description have essentially different extensions to two types of realization of the Poincar\`{e} group representations discussed by us.

\begin{center}
{\large \bf 2. Poincar\`{e} group representations of the standard type}
\end{center}

\begin{center}
{\bf 2.1. Two-fold representations of the rotation group in the quantum theory}
\end{center}

According to the group-theoretical definition, two representations $T_{1}(g)$ and $T_{2}(g)$ of the group $G$ realized in the linear spaces ${\cal L}_{1}$ and ${\cal L}_{2}$, accordingly, are equivalent if there exists a biunique linear correspondence between the vectors of these spaces kept unchanged under acting on them by operators $T_{1}(g)$ and $T_{2}(g)$ with any $g \in G$. It is supposed then, that the representations equivalent among themselves are not essentially different (see, for example, \cite{15}, \cite{16}). This agreement accepted in the group theory, strictly speaking, contradicts the physical reality, at least reflected by the quantum mechanics. So, since the times of formation of the quantum theory of hydrogen atom, three essentially different realizations of the rotational momentum of a particle are being considered, namely, the orbital, spin, and total angular momenta connected with both nonequivalent and mathematically equivalent representations of the spatial-rotation group $SO(3)$.

The components of the orbital angular momentum vector are associated with generators of the $SO(3)$-group representation acting in the space 
${\cal L}_{f}$ of the infinitely differentiable functions of spatial coordinates and having in that space the form
\begin{equation}
M^{\alpha \beta} = i(x^{\alpha}\partial/\partial x_{\beta}
-x^{\beta}\partial/\partial x_{\alpha}),
\label{1}
\end{equation}
where $\alpha, \beta = 1,2,3$. Then, in the space of irreducible representation characterized by the integer value of the orbital momentum $l$, the spherical functions $Y_{lm}(\theta, \varphi)$, where $m$ is the projection of the orbital momentum onto the third axis, are taken as the canonical basis In such a basis, a matrix form can be given to the generators $M^{\alpha\beta}$, however, with observing accuracy in notations and wording in order to not confuse the orbital angular momentum with the spin. The orbital angular momentum of a particle refers to its motion with respect to the fixed origin of coordinate space which often associates with the position of an atomic nucleus or with the interation point of particles. At some choice of the origin of coordinates the orbital angular momentum of a given particle may correspond to an irreducible representation of the group $SO(3)$, but at a different choice the representation generated by generators (\ref{1}) may change and become reducible.

The components of the spin vector are associated with the generators $S^{\alpha\beta}$ of the $SO(3)$-group representation, having, unlike the generators 
$M^{\alpha\beta}$, matrix realization in some linear space ${\cal L}_{a}$. The commutation relations for the family of the generators $M^{\alpha\beta}$ and for the family of the generators $S^{\alpha\beta}$ are identical. All pairs of generators from the different families commute among themselves. The space ${\cal L}_{a}$ is specified by an abstract (spin) canonical basis $\xi_{jk}$ with integer or half-integer values of the spin $j$ and its projection $k$ onto the third axis. The irreducible representations in ${\cal L}_{a}$ with integer spin values are mathematically equivalent to irreducible representations in ${\cal L}_{f}$ with approptiate integer values of the orbital angular momentum, but the irreducible representations in ${\cal L}_{a}$ with half-integer values of the spin have no analogue in ${\cal L}_{f}$. The spin is an internal characteristic of a given particle and does not depend on the choice of the origin of coordinates.

As a result, in the quantum mechanics, under a rotation of the spatial axes described by parameters $\epsilon_{\alpha \beta}$ ($\epsilon_{\alpha \beta}=-\epsilon_{\beta \alpha}$) and by an appropriate element $g \in SO(3)$, the wave function of a particle with nonzero spin transforms in accordance with a 
two-fold representation of the rotation group, namely, in accordance with the tensor product of two representations whose operators 
\begin{equation} 
T(g) = \exp (-\frac{i}{2} S^{\alpha \beta}\epsilon_{\alpha \beta}) \otimes 
\exp (-\frac{i}{2} M^{\alpha \beta}\epsilon_{\alpha \beta}) = 
\exp [-\frac{i}{2} (S^{\alpha \beta} \otimes E+E \otimes M^{\alpha \beta})\epsilon_{\alpha \beta}] 
\label{2}
\end{equation}
act in the tensor product of two linear spaces 
${\cal L}_{a}\otimes {\cal L}_{f}$, which we call the "hybrid" space.

If the representations of the rotation group in the spaces ${\cal L}_{a}$ and 
${\cal L}_{f}$ are irreducible and are described by nonzero values of the spin $j$ and the orbital angular momentum $l$, then the representation $T(g)$ 
(\ref {2}) is reducible and decomposes into a direct sum of irreducible representations, whereas the quantum numbers $J$ associated with the total angular momentum possess the values $|l-j|, |l-j|+1, \ldots, l+j$. Each of the canonical basic vectors $\zeta_{JM}$ representable in the form 
\begin{equation}
\zeta_{JJ_{3}} =\sum_{k+m=J_{3}} (j\;k\;l\;m|J\;J_{3})\xi_{jk}
Y_{lm}(\theta, \varphi),
\label{3}
\end{equation}
where $(j\;k\;l\;m|J\;J_{3})$ are the Clebsch-Gordon coefficients, contains a finite set of points from the function space ${\cal L}_{f}$ as well as points from the abstract space unequivocally associated with the former ones. Evidently, there is an infinite number of different realizations of irreducible representations of the group $SO(3)$ in the "hybrid" space ${\cal L}_{a}\otimes {\cal L}_{f}$, characterized by the same value of the total angular momentum 
$J$, but with different values of the orbital angular momentum and the spin.  According to the group-theoretical definition, these representations are equivalent to each other, though they refer to  a physically different internal structure, and lead to a mathematically different functional dependence on the spatial coordinates.

\begin{center}
{\bf 2.2. On some aspects of the description of the proper Lorentz group representations}
\end{center}

When constructing realizations of the proper Lorentz group representations, and, further on, the Poincar\`{e} group representations, we, after Gelfand and Yaglom \cite{12}, adhere to the statements formulated for the representations of the group of rotations in the Euclidean three-dimensional space and naturally extended to transformations in the Minkowski four-dimensional space-time.

As the proper Lorentz group $L^{\uparrow}_{+}$ is a group of transformations of the spatial coordinates and time, the dependence of the field of this or that particle on the space-time coordinates generates by itself some representation of it. The presence of a particle's spin requires that the field be also transformed under an $L^{\uparrow}_{+}$-group representation realized in some abstract (spin) space. Consequently, unless the particle field $\psi(x)$ is not a Lorentz scalar, it will be transformed, in all its states, as the tensor product of two $L^{\uparrow}_{+}$-group representations realized in the tensor product of the space of functions of the space-time coordinates ${\cal L}_{f}$ and of some abstract (spin) space ${\cal L}_{a}$.

Let a transition from one inertial reference frame to another one corresponds to an element $g$ of the group $L^{\uparrow}_{+}$ and to a transformation of the space-time coordinates $x = \{ x^{\mu}: \mu = 0,1,2,3 \}$ of the form
\begin{equation}
x' = \Lambda(g) x, \quad x'^{\mu} = [\Lambda(g)]^{\mu}{}_{\nu} x^{\nu},
\label{4}
\end{equation}
with
\begin{equation}
[\Lambda(g)]^{\mu}{}_{\nu}[\Lambda(g)]_{\mu}{}^{\rho} =
[\Lambda(g)]^{\rho}{}_{\mu}[\Lambda(g)]_{\nu}{}^{\mu} =
\delta_{\nu}^{\rho}.
\label{5}
\end{equation}
Then the proper Lorentz group representation in the space ${\cal L}_{f}$ is well-defined and is given by the formula
\begin{equation}
\varphi'(x) = \varphi(\Lambda^{-1}(g)x),
\label{6}
\end{equation}
where $\varphi(x)$ is a vector of this space. The expression for the generators 
$M^{\mu\nu}$ of representation (\ref{6}) results from expression (\ref{1}) by replacing the indexes $\alpha$, $\beta$ with $\mu$, $\nu$.

In the transformations of any vector $u$ from the abstract (spin) space 
${\cal L}_{a}$ corresponding to the $L^{\uparrow}_{+}$-group representation
\begin{equation}
u' = S(g)u,
\label{7}
\end{equation}
the operators $S(g)$ have matrix form, as well as the appropriate generators 
$S^{\mu\nu}$ of such a representation.

Every generator $S^{\mu\nu}$ commutes with every generator $M^{\rho\sigma}$. As a result, the transformation of a field $\psi(x)$ corresponding to a proper Lorentz group element $g$ realises in accordance with the tensor product of two representations, with operator $T(g)$ being similar to operator (\ref{2}). The field $\psi(x)$ serves as a vector of the "hybrid" space, namely, a vector of the tensor product of two linear spaces ${\cal L}_{a} \otimes {\cal L}_{f}$, and, due to relations (\ref{6}) and (\ref{7}), its Lorentz transformation has the following form
\begin{equation}
\psi'(x) = S(g) \psi (\Lambda^{-1}(g)x).
\label{8}
\end{equation}

The description of all irreducible representations of the group 
$L^{\uparrow}_{+}$ in the space ${\cal L}_{a}$ is given in paper \cite{12} (and in monographs \cite{15}, \cite{16}). It is obtained via solving the system of algebraic equations, which antisymmetric generators $S^{\mu\nu}$ of the proper Lorentz group obey,
\begin{equation}
[S^{\mu\nu}, S^{\rho\sigma}] = i(-g^{\mu\rho}S^{\nu\sigma}+ g^{\mu\sigma}
S^{\nu\rho} + g^{\nu\rho}S^{\mu\sigma} - g^{\nu\sigma}S^{\mu\rho}),
\label{9}
\end{equation}
where $g^{00}=-g^{11}=-g^{22}=-g^{33}=1$, $g^{\mu\nu}=0$ ($\mu \neq \nu$). Here, no requirements on the space ${\cal L}_{a}$ are imposed initially, except for the linearity: neither to be the Hilbert space, nor to be normalizable one.
 
Let us note in brief the aspects of the description of the proper Lorentz group representations which are the constitutive elements of our realization of the
Poincar\`{e} group representations and which have no analogues neither in the works with participation of Wigner \cite{8}, \cite{9} nor in the comprehensive monograph \cite {11}, while fragments of them are given in the monograph 
\cite{10}, already after discussion of the Wigner's realization of the 
Poincar\`{e} group representations.

In \cite{12}, the characterization of irreducible representations of the group 
$L^{\uparrow}_{+}$ begins with the list of numbers describing them and with  choosing the canonical basis in the space ${\cal L}_{a}$ that is connected with the rotation subgroup $SO(3)$. Thereby, the decomposition of such representations into a direct sum of irreducible representations of the group 
$SO(3)$ is specified. The results of action of the proper Lorentz group generators on the vectors of canonical basis are presented.

Really, each irreducible representation $\tau$ of the proper Lorentz group is defined by a pair of numbers $(j_{0},j_{1})$, where $2j_{0}$ is an integer and  $j_{1}$ is an arbitrary complex number. The canonical basis vectors of this representation space are denoted by $\xi_{\tau jk}$, where $j$ is the spin, $k$
is its projection onto the third axis, $k=-j,-j+1, \ldots, j$, and $j=|j_{0}|, 
|j_{0}|+1, \ldots$. The representation $\tau=(j_{0},j_{1})$ is 
finite-dimensional if $2j_{1}$ is an integer of the same parity as $2j_{0}$ and if $|j_{1}| > |j_{0}|$; then $j=|j_{0}|, |j_{0}|+1, \ldots, |j_{1}|-1$. The pairs  $(j_{0},j_{1})$ and $(-j_{0},-j_{1})$ define the same representation, $(j_{0},j_{1}) \sim (-j_{0},-j_{1})$. The formulas
\begin{eqnarray}
& &S^{12}\xi_{\tau jk} = k \xi_{\tau jk}, \label{10} \\
& &(S^{23}-iS^{31})\xi_{\tau jk} = \sqrt{(j-k+1)(j+k)} \xi_{\tau jk-1}, 
\label{11} \\
& &(S^{23}+iS^{31})\xi_{\tau jk} = \sqrt{(j-k)(j+k+1)} \xi_{\tau jk+1}, 
\label{12} \\
& &S^{30}\xi_{\tau jk} = B_{\tau j} \sqrt{j^{2}-k^{2}} \xi_{\tau j-1 k}
- A_{\tau j} k \xi_{\tau jk}- B_{\tau j+1} \sqrt{(j+1)^{2}-k^{2}} 
\xi_{\tau j+1 k}, \label{13} \\  
& &(S^{10}-iS^{20})\xi_{\tau jk} = -B_{\tau j} \sqrt{(j+k)(j+k-1)} 
\xi_{\tau j-1 k-1} \nonumber \\
& &\hspace{0.5cm} - A_{\tau j} \sqrt{(j+k)(j-k+1)} \xi_{\tau j k-1}
- B_{\tau j+1} \sqrt{(j-k+1)(j-k+2)} \xi_{\tau j+1 k-1},
\label{14} \\
& &(S^{10}+iS^{20})\xi_{\tau jk} = B_{\tau j} \sqrt{(j-k)(j-k-1)}
\xi_{\tau j-1 k+1} \nonumber \\
& &\hspace{0.5cm} - A_{\tau j} \sqrt{(j-k)(j+k+1)} \xi_{\tau j k+1}
+ B_{\tau j+1} \sqrt{(j+k+1)(j+k+2)} \xi_{\tau j+1 k+1}
\label{15} 
\end{eqnarray}
hold, where
\begin{equation}
A_{\tau j} = \frac{ij_{0}j_{1}}{j(j+1)}, \quad
B_{\tau j} = \frac{i}{j}
\sqrt{\frac{(j^{2}-j_{0}^{2})(j^{2}-j_{1}^{2})}{4j^{2}-1}}.
\label{16}
\end{equation}

Note, that only one nonzero value of spin $j$ is inherent in two nonequivalent irreducible representations $(j, j+1)$ and $(-j, j+1)$ and that one representation $(0,1)$ possesses only zero value of spin. An infinite number of other irreducible representations of the proper Lorentz group have not less than two values of spin.

After obtaining formulas (\ref{9})--(\ref{16}), the question of the existence and of the structure of the nondegenerate relativistic-invariant Hermitian bilinear form in the space ${\cal L}_{a}$ of this or that proper Lorentz group representation has been solved in \cite{12}. For two vectors 
$u_{n}=\sum_{\tau jk} w_{n\tau jk}\xi_{\tau jk}$, $n=1,2$, from the space 
${\cal L}_{a}$ this form reduces to
\begin{equation}
(u_{2}, \; u_{1})=\sum_{ \tau jk} w_{2\tau^{*} jk}^{*} a_{\tau^{*}\tau}(j) 
w_{1\tau jk},
\label{17}
\end{equation}
where $\tau^{*} = (-j_{0}, j_{1}^{*})$, if $\tau = (j_{0},j_{1})$, and the dependence of the quantity $a_{\tau^{*}\tau}(j)$ on spin $j$ is given by the recursive relation 
\begin{equation}
a_{\tau^{*}\tau}(j)=-\frac{B^{*}_{\tau^{*} j}}{B_{\tau j}}a_{\tau^{*}\tau}(j-1).
\label{18}
\end{equation}

It is follows from here, that an irreducible representation $\tau$ of the proper Lorentz group is unitary if it is equivalent to the representation 
$\tau^{*}$ and if the quantity $a_{\tau^{*}\tau}(j)$ does not depend on $j$. These conditions are fulfiled for two sets of representations $(j_{0},j_{1})$: (1) if the number $j_{1}$ is pure imaginary at any integer or half-integer value of $j_{0}$ (the principal series) or (2) if $j_{0}=0$, and $j_{1}$ is real, and $|j_{1}| \leq 1$ (the complementary series). Among the unitary irreducible representations of the group $L^{\uparrow}_{+}$, only one, namely the scalar $(0, 1)$, is finite-dimentional, all other are infinite-dimentional.

The question of a possible realization of infinite-dimentional irreducible representations of the proper Lorentz group in the space of functions of the space-time coordinates ${\cal L}_{f}$, most likely, was considered by nobody. It is easy to show that, from all sets of finite-dimentional irreducible representations of the group $L^{\uparrow}_{+}$, only the representations 
$(0, n)$, $n=1,2,\ldots$, admit the realization in ${\cal L}_{f}$. The representation $(0, n)$ contains orbital angular momenta $0,1,\ldots, n-1$, and the space corresponding to it, ${\cal L}_{f}^{(0,n)}$, consists of homogeneous polynoms on coordinates $x^{\mu}$ of degree $n-1$. This space can be considered as a linear hull spanned by the components 
$\chi^{\mu_{1}\ldots \mu_{n-1}}_{(0,n)}$ of appropriate totally symmetric traceless tensor of the rank $n-1$. For example, such tensors of the lowest ranks have the following form: $\chi_{(0,1)} = 1$; 
$\chi^{\mu}_{(0,2)} = x^{\mu}$; $\chi^{\mu_{1}\mu_{2}}_{(0,3)} = 
x^{\mu_{1}}x^{\mu_{2}}-(1/4) x^{2}g^{\mu_{1}\mu_{2}}$; 
$\chi^{\mu_{1}\mu_{2}mu_{3}}_{(0,4)} = x^{\mu_{1}}x^{\mu_{2}}x^{\mu_{3}}- 
(1/6)x^{2}[g^{\mu_{1}\mu_{2}} x^{\mu_{3}}+g^{\mu_{1}\mu_{3}} x^{\mu_{2}}+
g^{\mu_{2}\mu_{3}} x^{\mu_{1}}]$.

In the field theory and in particle physics, there was no urgent need in introducing a canonical basis in the space ${\cal L}_{f}^{(0,n)}$ and in  finding an analogue of relation (\ref{3}) for the irreducible representations  in the "hybrid" space ${\cal L}_{a} \otimes {\cal L}_{f}$ when realizing in it 
the tensor product $(j_{0}, j_{1}) \otimes (0, n)$ of all irreducible representations of the proper Lorentz group admissible in the corresponding spaces.

\begin{center}
{\bf 2.3. Preliminary remarks on the irreducible representations of the 
Poincar\`{e} group in various spaces}
\end{center}

As it is known, the proper (or orthochronous) Poincar\`{e} group is a semidirect product of the proper (or orthochronous) Lorentz group and the 
four-parametrical abelian group of space-time translations $T_{4}$ playing the role of invariant subgroup. The translation group $T_{4}$ is generated by the transformations $x'^{\mu} = x^{\mu}+a^{\mu}$, and the generators of its representations in the function space ${\cal L}_{f}$ are identified with the operator of the momentum $P^{\mu}$, which is a multiple of the differentiation operator: $P^{\mu} = i \partial / \partial x_{\mu}$. The irreducible representations of the translation group $T_{4}$ in the space of differentiable limited functions of the space-time coordinates are one-dimentional. Each of them is associated with a basic vector of the type $\exp (-iq^{\mu}x_{\mu})$ with a fixed value of the four-momentum $q^{\mu}$. This or that transformation of the proper Lorentz group, according to formulas (\ref{4}) and (\ref{6}), convertes such a basic vector to another one. Thereof, an orbit in the space ${\cal L}_{f}$ generated by all Poincar\`{e} group transformations and containing the vector $\exp (-iq^{\mu}x_{\mu})$ consists of vectors of the type $c\exp (-ip^{\mu}x_{\mu})$ with arbitrary numerical phase factors $c$ and with any of four-vectors $p^{\mu}$ satisfying the condition $p^{2}=q^{2}$. The linear space spanned by the vectors of such an orbit is the space of the irredusible Poincar\`{e} group representation characterized by a well-defined proper value of the momentum operator squared.

Since the variations in the choice of the coordinate origin and the time reference point cannot affect the spin as internal characteristic of a particle, then, in the abstract (spin) space ${\cal L}_{a}$, the identical transformation is set to correspond to the translation group $T_{4}$, so that any Poincar\`{e} group representation in this space is nothing else but the proper Lorentz group representation. Therefore, the irredusible Poincar\`{e} group representations in the function space ${\cal L}_{f}$ and in the abstract space ${\cal L}_{a}$ have essentially different characteristics.

In the standard approach which we discuss here, the description of Poincar\`{e} group representations potentially suitable for the field theory referes to the "hybrid" space ${\cal L}_{a} \otimes {\cal L}_{f}$ and is specified by the operator
\begin{equation} 
T(g) =  \exp [-i(E \otimes P^{\mu}) a_{\mu}-\frac{i}{2} (S^{\mu \nu} \otimes E
+E \otimes M^{\mu \nu})\epsilon_{\mu \nu}]. 
\label{19}
\end{equation}

Using the recipes of section 2.5 below-mentioned, we could construct some set of irredusible Poincar\`{e} group representations in the "hybrid" space, every time starting from a pair of basic vectors with any admissible values of their indices, $\xi_{(j_{0}, j_{1})jk}$ and $\exp (-iq^{\mu}x_{\mu})$, from the representation spaces ${\cal L}_{a}$ and ${\cal L}_{f}$ accordingly. Then, most likely, it would appear impossible to relate the representations of such a set to any linear relativistic-invariant equations. Therefore, directing our efforts to the field theory and particle physics, we find it reasonable to pay main attention to constructing the Poincar\`{e} group representations in the spaces of solutions of such equations and to considering the question of their unitarity and irredusibility.

\begin{center}
{\bf 2.4. Some assertions about the linear relativistic-invariant equations}
\end{center}

We note, first of all, the difference and the similarity in the definition of four-vector operators of the proper Lorentz group in the space of functions  
${\cal L}_{f}$ and in the abstract (spin) space ${\cal L}_{a}$.

In the space of functions of the space-time coordinates, it is admissible to refer to the four-vector operators $V^{\mu}$ what are represented in the form 
\begin{equation}
V^{\mu} = F_{1}\left( x^{\nu}x_{\nu}, x^{\nu}\frac{\partial}{\partial x^{\nu}},
\frac{\partial^{2}}{\partial x_{\nu}\partial x^{\nu}} \right) x^{\mu}
+F_{2}\left( x^{\nu}x_{\nu}, x^{\nu}\frac{\partial}{\partial x^{\nu}},
\frac{\partial^{2}}{\partial x_{\nu}\partial x^{\nu}} \right)
\frac{\partial}{\partial x_{\mu}},
\label{20}
\end{equation}
where $F_{1}$ and $F_{2}$ are arbitrary functions of their arguments. 
At the transition from one inertial reference frame to another, all operators 
$V^{\mu}$, due to relation (\ref{4}), transform just as four-vectors, namely, under the formula
\begin{equation}
V'^{\mu} = \Lambda^{\mu}{}_{\nu}(g) V^{\nu},
\label{21}
\end{equation}
which does not depend on the proper Lorentz group representation realized in the function space. At the same time, the four-vector operators $V^{\mu}$ satisfy the commutation relations of the form
\begin{equation}
[M^{\mu\nu}, V^{\rho}] = i(-g^{\mu\rho} V^{\nu}+ g^{\nu\rho} V^{\mu}),
\label{22}
\end{equation}
as it follows from the formula of the type (\ref{1}) for the generators 
$M^{\mu\nu}$ and from formula (\ref{20}).

We are guided by the following definition concerning the matrix operators 
$\Gamma^{\mu}$ and $R$ acting in an abstract (spin) space ${\cal L}_{a}$ of some $L^{\uparrow}_{+}$-group representation: they are, respectively, the
four-vector and scalar operators, if, at any choice of a four-vector 
$\eta^{\mu}$, the bilinear forms $(u_{2}, \Gamma^{\mu}\eta_{\mu}u_{1})$ and 
$(u_{2}, Ru_{1})$, where $u_{1}, u_{2} \in {\cal L}_{a}$, are relativistically invariant. It is follows from here and from (\ref{7}) and (\ref{21}), that the matrix operators $\Gamma^{\mu}$ and $R$ should obey conditions
\begin{equation}
S^{-1}(g) \Gamma^{\mu} \Lambda_{\mu}{}^{\nu}(g) S(g) = \Gamma^{\nu}, 
\label{23}
\end{equation}
\begin{equation}
S^{-1}(g) R S(g) = R 
\label{24}
\end{equation}
and the commutation relations, resulting from (\ref{23}) and (\ref{24}),
\begin{equation}
[S^{\mu\nu}, \Gamma^{\rho}] = i(-g^{\mu\rho} \Gamma^{\nu}
+ g^{\nu\rho} \Gamma^{\mu}),
\label{25}
\end{equation}
\begin{equation}
[S^{\mu\nu}, R] = 0.
\label{26}
\end{equation}

While the condition (\ref{23}), unlike (\ref{21}), obviously depends on the proper Lorentz group representation realized in the corresponding space, the commutation relations (\ref{25}) and (\ref{22}) for the four-vector operators in physically different spaces are identical to each other.

The requirement of the relativistic invariance (invariance under the orthochronous Lorentz group transformations) of the linear equation for a field $\psi(x)$
\begin{equation}
(\Gamma^{\mu} \frac{\partial}{\partial x^{\mu}} + i R) \psi (x) = 0
\label{27}
\end{equation} 
translates by virtue of relations (\ref {8}) and (\ref {21}) to conditions 
(\ref {23}) and (\ref {24}) to which the matrix operators $\Gamma^{\mu}$ and 
$R$ entering the equation should subject.

Note now some general details. The equation (\ref{27}) is initially specified in the "hybrid" space of the proper Lorentz group representation. It is constructed on the contraction of the four-vector operators $\partial_{\mu}$ and $\Gamma^{\mu}$, acting in different spaces, accordingly, in ${\cal L}_{f}$ and ${\cal L}_{a}$. It is easy to make sure using relations (\ref{22}) and 
(\ref{23}) that this contraction commutes with the generator $S^{\nu\rho} \otimes E+E \otimes M^{\nu\rho}$ of the $L^{\uparrow}_{+}$-group  representation in the space ${\cal L}_{f} \otimes{\cal L}_{a}$, i.e. it is a scalar in this space. The structure of equation (\ref {27}) only depends on the set of irreducible representations contained in the matrix representation 
$S(g)$. The choice of the latter is each time made by us on the basis of those or other reasons. The proper Lorentz group representations in the space of functions ${\cal L}_{f}$, which are assigned to a field $\psi (x)$, are never specified. They are completely defined in an implicit form by concrete solutions of equation (\ref{27}). This equation possesses also the invariance under the space-time translations. Thereby it is invariant under the Poincar\`{e} group transformations.

Being guided by applications in particle physics, we exclude from our consideration the FSIIR-class fields which transform under the representations 
$S(g)$ of the group $L^{\uparrow}_{+}$ decomposable into a finite direct sum of infinite-dimentional irreducible representations, because the corresponding equations (\ref{27}) possess mass spectra with an accumulation point at zero \cite{12}, \cite{15}, \cite{20}, \cite{21}, and they also can possess spacelike (tachyonic) solutions \cite{22}. Besides that, the FSIIR-class field theory can have such "deceases" as the lack of $CPT$-invariance \cite{23}, the violation of conventional connection between spin and statistics \cite{24}, the local noncommutativity of fields \cite {25}. In what follows, we restrict ourselves to fields of class ISFIR or class FSFIR only, whose representations $S(g)$ are decomposable into an infinite or finite direct sum of finite-dimentional irreducible representations of the proper Lorentz group, and which theory does not suffer from "deceases" listed above and does not possess tachyonic states.

On the form of equation (\ref{27}), we have the right to believe that it may have solutions in the form of flat waves, which correspond to the definite values of the four-momentum $p^{\mu}$ of $\psi(x)$-field states. For every field state with a timelike four-momentum ($p^{2}=M^{2} > 0$), we introduce the rest system in which ${\bf p} = 0$, i.e. we seek solutions of equation 
(\ref{27}) in the form $\psi_{0}(x) = u(p_{0})\exp(-iMt)$. Then, the vectors 
$u(p_{0})$ from the space ${\cal L}_{a}$ should obey the following equation
\begin{equation}
(\Gamma^{0} M - R) u(p_{0}) = 0.
\label{28}
\end{equation} 

The general solution of the system (\ref {25}) with respect to the 
four-vector operators $\Gamma^{\mu}$ has been found in \cite{12} via using relations (\ref{9})--(\ref{16})
\begin{eqnarray}
\Gamma^{0} \xi_{(j_{0}, j_{1}) jk} &=&
c(j_{0}+1,j_{1};j_{0},j_{1})
\sqrt{(j+j_{0}+1)(j-j_{0})} \xi_{(j_{0}+1, j_{1}) jk} \nonumber \\
&+& c(j_{0}-1,j_{1};j_{0},j_{1}) \sqrt{(j+j_{0})(j-j_{0}+1)}
\xi_{(j_{0}-1, j_{1}) jk} \nonumber \\
&+& c(j_{0},j_{1}+1;j_{0},j_{1}) \sqrt{(j+j_{1}+1)(j-j_{1})}
\xi_{(j_{0}, j_{1}+1) jk} \nonumber \\
&+& c(j_{0},j_{1}-1;j_{0},j_{1}) \sqrt{(j+j_{1})(j-j_{1}+1)}
\xi_{(j_{0}, j_{1}-1) jk},
\label{29}
\end{eqnarray}
where $c(j'_{0},j'_{1};j_{0},j_{1})$ are arbitrary, and
\begin{equation}
\Gamma^{n} = -i[S^{n0}, \Gamma^{0}],
\label{30}
\end{equation}
with $n=1,2,3$.

From relation (\ref{26}), we have
\begin{equation}
R \xi_{(j_{0}, j_{1}) jk} = r(j_{0}, j_{1}) \xi_{(j_{0}, j_{1}) jk},
\label{31}
\end{equation}
where $r(j_{0}, j_{1})$ are arbitrary.

The requirement that equation (\ref{27}) be invariant under the spatial reflection and the reality condition for the Lagrangian for free field 
$\psi(x)$ corresponding to equation (\ref{27}) impose some restrictions on the sets of constants $c(j'_{0},j'_{1};j_{0},j_{1})$ and $r(j_{0}, j_{1})$ (see \cite{12}, \cite{15}). It is easy to make sure that in the space of fields of class ISFIR or class FSFIR these restrictions provide the hermicity of the operator $\Gamma^{0}$ and then, due to relations (\ref{30}) and (\ref{13})--(\ref{16}), the antihermicity of the operators $\Gamma^{n}$, $n=1,2,3$.

Formulas (\ref{29}) and (\ref{30}) show that in the general case the four-vector operator $\Gamma^{\mu}$ couples the given irreducible representation of the proper Lorentz group  with four other irreducible representations. There are only two irreducible representations of the group $L^{\uparrow}_{+}$, namely, the infinite-dimentional representations $(0, 1/2)$ and $(1/2, 0)$, named Majorana representations, for each of which (due to their equivalence to the representations $(0, -1/2)$ and $(-1/2, 0)$ correspondingly) the operator 
$\Gamma^{\mu}$ can couple with itself. Except for Dirac equation based on the representation $(-1/2, 3/2) \otimes (1/2, 3/2)$ and describing states with only one value of spin, all other equations (\ref{27}) lead to a set of states, with which not less than two values of spin are associated.

Note that the matrix elements of the operator $\Gamma^{0}$ given by relation 
(\ref{29}) are diagonal in spin $j$ and in its projection $k$, and are independent of $k$. It results from that the operator $\Gamma^{0}$ commutes with the rotation group generators. In turn, this follows from that the contraction of the four-vector operator $\Gamma^{\mu}$ with any four-vector 
$\eta^{\mu}$ or with the four-vector operator $\partial^{\mu}$ is the scalar operator in the space ${\cal L}_{a} \otimes {\cal L}_{f}$ and in such a contraction the operator $\Gamma^{0}$ is coupled with the time component 
$\eta^{0}$ or with the time derivative which do not change under spatial rotations.

Relations (\ref{29}) and (\ref{31}) result in splitting equation (\ref{28}) into a system of independent equations, each of which is characterized by some values of spin $j$ and its projection $k$, by the set of masses, identical to all spin projections, and by its state vectors $u(p_{0})$. Those values of the quantity $M$, at which the nontrivial normalized solutions for vectors 
$u(p_{0})$ do exist, are assigned to the mass spectrum.

The requirement of normalizability of these vectors is only necessary when the proper Lorentz group representation $S(g)$ assigned to a field $\psi (x)$ from equation (\ref{27}) is decomposable into a infinite direct sum of 
finite-dimentional irreducible representations. It can be expressed either as formal mathematical requirement of finiteness of the relativistically invariant bilinear form $(u(p_{0}), \; u(p_{0}))$, or as physical requirement of finiteness of the amplitudes of various processes involving the corresponding particles, as it was proposed in \cite{19}.

\begin{center}
{\bf 2.5. Orbits and linear hulls generated by solutions of the relativistic-invariant equations}
\end{center}

Each mass value $M$ and the corresponding vector $u(p_{0})$ found from equation 
(\ref{28}) is associated with some particle. As the state vectors of different particles are realized in the initial representation space assigned to equation (\ref{27}), then, when studying a given particle, we should concentrate on the orbits of its states, which represent the manifolds of vectors in the representation space obtained from the vectors of the rest state by applying all possible proper Lorentz group transformations, and on the linear hulls spanned by such orbits.

If the quantity $M$ from equation (\ref{28}) is a point of the mass spectrum at some value of spin $j$ and any value of its projection $k$, then the solution of this equation with respect to vector components $u_{jk}(p_{0})$ in the canonical basis of $S(g)$-representation space gives for the field 
$\psi_{0jk}(x)$ in the rest system of the corresponding particle the unambiguous (up to an unessential factor) expression
\begin{equation}
\psi_{0jk}(x) = u_{jk}(p_{0}) \exp(-ip_{0}x) = \sum_{\tau \in S(g)} 
w_{M\tau j} \xi_{\tau jk} \exp(-ip_{0}x),
\label{32}
\end{equation}
where $p_{0} = \{M, 0, 0, 0 \}$. 

As it is known (see, for example, \cite{8} \cite {15}), a transition from the initial inertial reference frame to any other inertial  reference frame may be carried out by three operations: by some rotation of the coordinate system, by transition to the reference frame moving relative to the initial one along its third axis with the velocity $v$ (by a boost), and by another rotation of the coordinate system. As the finite transformations of the rotation group $SO(3)$ are well-known, it is possible to restrict our consideration of the Lorentz transformations of a field to boosts along the third axis. We have from the equality (\ref{32})
\begin{eqnarray}
& &\psi_{\alpha jk}(x) = \exp (iS^{30}\alpha) u_{jk}(p_{0}) \exp(-ip_{\alpha}x) \nonumber \\
& &=\sum_{(j_{0},j_{1})\in S(g)} \sum_{j'\geq |j_{0}|} 
 A^{(j_{0},j_{1})}_{j'k,jk}(\alpha) w_{M(j_{0}, j_{1}) j}
\xi_{(j_{0}, j_{1}) j'k} \exp(-ip_{\alpha}x),
\label{33}
\end{eqnarray}
where $\tanh \alpha = v$, $p_{\alpha}=\{ M\cosh\alpha, 0, 0, M\sinh\alpha \}$,
and $A^{(j_{0},j_{1})}_{j'k,jk}(\alpha)$ are the matrix elements of the operator
$\exp (iS^{30}\alpha)$ of the finite Lorentz transformations, with 
$A^{(j_{0},j_{1})}_{j'k,jk}(0) = \delta_{j'j}$. It follows from here, that every non-Dirac particle at a transition from the rest frame where it has the definite spin $j$ to any other inertial reference frame takes all values of spin, which are inherent in the proper Lorentz group representation $S(g)$ under consideration.
 
The problem of explicit form of matrix elements 
$A^{(j_{0},j_{1})}_{j'k,jk}(\alpha)$ found via using relations 
(\ref {13}), (\ref {16}) is solved far not to the full measure. Some set of such elements is obtained in \cite{26} for the unitary irreducible representations of the group $L^{\uparrow}_{+}$, and another one in \cite{27} for finite-dimentional irreducible representations. In particular, we have for the letter case
\begin{eqnarray}
& &A^{(\frac{1}{2},j_{1})}_{\frac{1}{2} \frac{1}{2},\frac{1}{2}\frac{1}{2}}(\alpha) 
= A^{(-\frac{1}{2},j_{1})}_{\frac{1}{2} \frac{1}{2},\frac{1}{2}\frac{1}{2}}
(-\alpha)
=A^{(\frac{1}{2},j_{1})}_{\frac{1}{2} -\frac{1}{2},\frac{1}{2}-\frac{1}{2}}
(-\alpha) \nonumber \\
& &=\frac{2}{j_{1}^{2}-1/4} \sum_{N=0}^{j_{1}-3/2} (j_{1}-N-1/2) 
\exp[(j_{1}-2N-1)\alpha].
\label{34}
\end{eqnarray}
As soon as this or that problem of particle physics demands knowing a new set of matrix elements of the operator $\exp (iS^{30}\alpha)$, they, undoubtedly, will be found.

Let us consider an orbit in the space ${\cal L}_{a} \otimes {\cal L}_{f}$, which is generated by all transformations of the proper Lorentz group acting on the state vector $\psi_{0}(x)$ of a rest particle with mass $M$, spin $j$ and its projection $k$
\begin{equation}
\psi_{gjk}(x) = S(g) u_{jk}(p_{0}) \exp(-ip_{g}x),
\label{35}
\end{equation}
where $p_{g} = \Lambda(g) p_{0}$, $g \in L^{\uparrow}_{+}$. Each vector from orbit (\ref{35}) is a solution of the relativistic-invariant equation 
(\ref{27}) and possesses the definite value of the four-momentum $p_{g}$, but the same value of the momentum $p_{g}$ belongs to an infinite set of vectors from this orbit assigned to various orientations of the coordinate axes of some inertial reference frame.

We form now a linear hull ${\cal L}_{jp_{0}}$, which consists of solutions of equation (\ref{27}) and is spanned by orbit (\ref{35}). As it follows from our construction, the linear hull ${\cal L}_{jp_{0}}$ is invariant under the proper Lorentz group transformations and under the translations of the space-time coordinates. 

\begin{center}
{\bf 2.6. An infinite number of Poincar\`{e} group representations characterized by the same proper values of the Casimir operators}
\end{center}

Let's prove now, that each vector of the linear hull ${\cal L}_{jp_{0}}$ spanned by orbits (\ref{35}) is a proper vector of the Casimir operators of the 
Poincar\`{e} group with the same proper values expressed through the mass $M$ and the rest spin $j$ of a particle.

As is known, it is possible to construct two and only two Casimir operators \cite{9}, \cite{11} from the generators of the Poincar\`{e} group representation acting in the space ${\cal L}_{a}\otimes {\cal L}_{f}$, namely, from $E\otimes P^{\mu}$ and  $L^{\mu \nu} \equiv S^{\mu \nu} \otimes E+E \otimes M^{\mu \nu}$. These are
\begin{equation} 
P = P^{\mu}P_{\mu}
\label{36}
\end{equation}
and
\begin{equation} 
W = (1/2) S^{\mu\nu}S_{\mu\nu}P^{\rho}P_{\rho} - 
S^{\mu\rho}S_{\nu\rho}P_{\mu}P^{\nu}.
\label{37}
\end{equation}
The Casimir operator $W$ represents the square of the Pauli-Lubanski four-vector
$W^{\mu}$, of which the operator $M^{\mu \nu}$ drops out because of its structure of the type (\ref{1}), 
\begin{equation} 
W^{\mu} = (1/2) \varepsilon^{\mu\nu\rho\sigma}L_{\nu\rho}(E\otimes P_{\sigma})
= (1/2) \varepsilon^{\mu\nu\rho\sigma}S_{\nu\rho} \otimes P_{\sigma}.
\label{38}
\end{equation}
 
Any vector (\ref{35}), and also, consequently, any vector of the linear hull 
${\cal L}_{jp_{0}}$ is evidently a proper vector of the Casimir operator $P$ with the proper value equal to the mass square $M^{2}$ of the respective particle.

In turn, acting by the Casimir operator $W$ on the vector of the particle rest state (\ref{32}) gives after the relations (\ref{10})--(\ref{12})
\begin{eqnarray}
& &W\psi_{0jk}(x) = \sum_{\tau \in S(g)} w_{M\tau j}
[(1/2) S^{\mu\nu}S_{\mu\nu}\cdot (p_{0})^{\rho}(p_{0})_{\rho}  \nonumber \\ 
& &- S^{\mu\rho}S_{\nu\rho}\cdot (p_{0})_{\mu}(p_{0})^{\nu}]
\xi_{\tau jk} \exp(-ip_{0}x) \nonumber \\
& &= M^{2} \sum_{\tau \in S(g)} w_{M\tau j}
\left[\sum_{\alpha, \beta = 1}^{3} (1/2) S^{\alpha \beta} S_{\alpha \beta}\right] \xi_{\tau jk} \exp(-ip_{0}x) \nonumber \\
& &= M^{2} j(j+1) \psi_{0}(x).
\label{39}
\end{eqnarray}
As the operator $W$ commutes with all generators $L^{\mu \nu}$ of the proper Lorentz group in the space ${\cal L}_{a} \otimes {\cal L}_{f}$, it also commutes with operators of transformations of the group $L^{\uparrow}_{+}$ translating the rest state vector $\psi_{0jk}(x)$ into orbit vectors $\psi_{gjk}(x)$ (\ref{35}). Consequently, the action of the Casimir operator $W$ on vector (\ref{35}) reduces to its action on the vector $\psi_{0jk}(x)$, i.e., by virtue of relation (\ref{39}), to multiplying any vector $\psi_{gjk}(x)$ (\ref{35}), and so, any vector of the linear hull ${\cal L}_{jp_{0}}$, by the same number $M^{2} j(j+1)$. In view of all importance of this conclusion, we present below a detailed consideration of how the operator $W$ (\ref{37}) acts on the vector 
$\psi_{gjk}(x)$ (\ref {35}).

Note in the beginning that the contraction of the proper Lorentz group generators $S^{\mu \nu}$ in the abstract (spin) space ${\cal L}_{a}$ and an antisymmetric tensor $\epsilon_{\mu \nu}$ leaves the bilinear form $(u_{2}, \; S^{\mu\nu}\epsilon_{\mu\nu}u_{1})$, where $u_{1}, u_{2} \in {\cal L}_{a}$, relativistically invariant. It is follows from here, that operators 
$S^{\mu \nu}$ obey the condition
\begin{equation}
S^{-1}(g) S^{\rho\sigma} S(g) {[U(g)]_{\rho\sigma}}^{\mu\nu} = S^{\mu\nu},
\label{40}
\end{equation}
if Lorentz transformation of the antisymmetric тензора $\epsilon_{\mu \nu}$ has the form
\begin{equation}
\epsilon'_{\mu\nu} ={[U(g)]_{\mu\nu}}^{\rho\sigma} \epsilon_{\rho\sigma},
\label{41}
\end{equation}
where, due to relations (\ref {5}) and (\ref {21}),
\begin{equation}
{[U(g)]_{\mu\nu}}^{\rho\sigma} = \frac{1}{2} \{ {[\Lambda(g)]_{\mu}}^{\rho}
{[\Lambda(g)]_{\nu}}^{\sigma} - {[\Lambda(g)]_{\nu}}^{\rho} 
{[\Lambda(g)]_{\mu}}^{\sigma} \} ,
\label{42}
\end{equation}
\begin{equation}
{[U(g)]_{\mu\nu}}^{\tau\xi} {[U(g)]^{\rho\sigma}}_{\tau\xi}= \frac{1}{2} 
(\delta_{\mu}^{\rho}\delta_{\nu}^{\sigma} -\delta_{\nu}^{\rho}
\delta_{\mu}^{\sigma}).
\label{43}
\end{equation}

Using relations (\ref{5}), (\ref{21}), (\ref{35}), (\ref{37}), (\ref{40}), 
(\ref{42}), and (\ref{43}), we obtain the following chain of equalities
\begin{eqnarray}
& &W\psi_{gjk}(x) = \{ (1/2) S^{\mu\nu}S_{\mu\nu}\cdot (p_{g})^{\rho}
(p_{g})_{\rho} - S^{\mu\rho}S_{\nu\rho}\cdot (p_{g})_{\mu}(p_{g})^{\nu} \} 
S(g) u(p_{0}) \exp(-ip_{g}x) \nonumber \\
& &=\{ (1/2) M^{2} S^{\mu\nu}S_{\rho\omega}  (1/2)[\delta_{\mu}^{\rho}
\delta_{\nu}^{\omega}-\delta_{\nu}^{\rho}\delta_{\mu}^{\omega}] \nonumber \\
& &- \delta_{\rho}^{\sigma}S^{\mu\rho}
S_{\nu\sigma}{[\Lambda(g)]_{\mu}}^{\tau} {[\Lambda(g)]^{\nu}}_{\omega}
(p_{0})_{\tau}(p_{0})^{\omega} \}S(g) u(p_{0}) \exp(-ip_{g}x) \nonumber \\
& &= \{ (1/2) M^{2} S^{\mu\nu}S_{\rho\omega}{[U(g)]_{\mu\nu}}^{\tau\xi}
 {[U(g)]^{\rho\omega}}_{\tau\xi}\nonumber \\
& &- {[\Lambda(g)]_{\rho}}^{\xi} {[\Lambda(g)]^{\sigma}}_{\xi}
S^{\mu\rho}S_{\nu\sigma}{[\Lambda(g)]_{\mu}}^{\tau} 
{[\Lambda(g)]^{\nu}}_{\omega}(p_{0})_{\tau}(p_{0})^{\omega} \}
S(g) u(p_{0}) \exp(-ip_{g}x) \nonumber \\
& &= \{ (1/2) M^{2} S^{\mu\nu}{[U(g)]_{\mu\nu}}^{\tau\xi}S_{\rho\omega} 
{[U(g)]^{\rho\omega}}_{\tau\xi} \nonumber \\
& &-S^{\mu\rho}{[U(g)]_{\mu\rho}}^{\tau\xi}
S_{\nu\sigma}{[U(g)]^{\nu\sigma}}_{\omega\xi}(p_{0})_{\tau}(p_{0})^{\omega} \}
S(g) u(p_{0}) \exp(-ip_{g}x) \nonumber \\
& &=\{ (1/2) M^{2} S(g)S^{\tau\xi}S^{-1}(g)S(g)S_{\tau\xi}S^{-1}(g)\nonumber \\
& &-S(g)S^{\tau\xi}S^{-1}(g)S(g)S_{\omega\xi}S^{-1}(g)
(p_{0})_{\tau}(p_{0})^{\omega} \}S(g) u(p_{0}) \exp(-ip_{g}x)\nonumber \\
& &= M^{2} S(g) \{ (1/2)S^{\tau\xi}S_{\tau\xi}-S^{0\xi}S_{0\xi} \} u(p_{0}) 
\exp(-ip_{g}x)\nonumber \\
& &= M^{2} S(g)\sum_{\tau \in S(g)} w_{M\tau j}
\left\{ \sum_{\alpha, \beta = 1}^{3} (1/2) S^{\alpha \beta} S_{\alpha \beta}
\right\} \xi_{\tau jk} \exp(-ip_{g}x)\nonumber \\
& &= M^{2} j(j+1) \psi_{gjk}(x).
\label{44}
\end{eqnarray}

So, each of the Casimir operators $P$ (\ref{36}) and $W$ (\ref{37}) of the 
Poincar\`{e} group in linear space, which is the linear hull 
${\cal L}_{jp_{0}}$ of orbits (\ref{35}), is proportional to the identity operator. Strictly speaking, it gives an evidence of the validity of only the necessary condition of the irreduciblity of the Poincar\`{e} group representation realized in space ${\cal L}_{jp_{0}}$. We will return to the question of irreducibility of such representations in the end of section 2.7.

In any case, we now have right to make positive conclusion on the existence of an infinite number of nonequivalent irreducible representations of the 
Poincar\`{e} group characterized by same proper values of the Casimir operators $P$ and $W$. Indeed, it is enough for this purpose to take into consideration, that there exist an infinite number of the proper Lorentz group representations $S(g)$ in the abstract (spin) space ${\cal L}_{a}$, containing the prescribed spin $j$, and that there exist an infinite number of relativistic-invariant equations (\ref{27}), one of the solutions of which describes the state 
(\ref {32}) of a particle with rest spin $j$ and generates the space ${\cal L}_{jp_{0}}$. Further, we are able to provide the same predefined value of the mass of the rest spin $j$ particle for each of the considered equations 
(\ref{27}) by the normalization of the four-vector operator $\Gamma^{\mu}$ or the scalar operator $R$.

Note, that the basis of a representation of the Poincar\`{e} group in the linear space ${\cal L}_{jp_{0}}$ remains unknown. At the same time, no need is seen in having it for further using in field-theoretical applications in particle physics.

\begin{center}
{\bf 2.7. On the unitarity and irreducibility of the Poincar\`{e} group representations in linear hulls ${\cal L}_{jp_{0}}$}
\end{center}

In his work \cite{8}, Wigner addresses to a quantum mechanical rule, according to which the probability of transition from one state to another, described in some inertial reference frame by the wave functions $\phi_{1}$ and $\phi_{2}$ respectively, is given by the square of the modulus of their scalar product 
$|(\phi_{2},\phi_{1})|^{2}$. As the mentioned probability should be the same in all inertial reference frames, then, according to Wigner's opinion, to describe a transition from one reference frame to another it is necessary to define the transformation of wave functions as a linear unitary operator. This opinion, containing both the definition and the requirement of unitarity of Lorenz transformations of particle fields, is not correct enough.

In \cite{15}, one finds the following definition: the unitarity of a representation of the group $G$ realized by operators $T(g)$ in the linear space ${\cal L}$ means that there exists a positively definite bilinear Hermitian form ($(\phi, \phi)> 0$ for every vector $\phi \in {\cal L}$), which is invariant under the action of operators $T(g)$ for all $g \in G$.

Comparing these two definitions of the unitarity of a representation we see that the question of positive definiteness of scalar product of wave functions does not arise in \cite{8}. Hence, the requirement stated by Wigner is only the requirement of relativistic invariance of the bilinear form, that does not at all mean the unitarity of transformations of wave vectors. At the same time in Wigner's construction of the Poincar\`{e} group representations, at least in the space of functions of the time-like four-momenta, the positive definiteness of the scalar product and the unitarity of transformations is provided automatically, no nonunitary representations which should be eliminated arise.

Note also, that, in the Lagrange's field theory, the kinetic and mass terms of the free Lagrangians and the field currents in the interaction Lagrangians are expressed through relativistically invariant bilinear forms, and in each of the above cases, the positive definiteness of these forms and, thereby, the unitarity of representations of the Lorentz or Poincar\`{e} group is not required.

Nevertheless, let's analyse now the situation with the values of relativistically invariant bilinear form in the linear hull ${\cal L}_{jp_{0}}$.

Using relation (\ref{8}), we conclude that the bilinear form 
$(\psi_{2}, \psi_{1})_{{\cal L}_{a} \otimes {\cal L}_{f}}$ for vectors
$\psi_{2}(x)$ and $\psi_{1} (x)$ from the space ${\cal L}_{a}\otimes 
{\cal L}_{f}$, expressed through relativistically invariant bilinear form 
(\ref{17}) for vectors from the space ${\cal L}_{a}$ by means of the following equality 
\begin{equation}
(\psi_{2}, \psi_{1})_{{\cal L}_{a} \otimes {\cal L}_{f}} 
= \int [(i\frac{\partial}{\partial t}\psi_{2}(x), \; \psi_{1} (x)) + 
(\psi_{2}(x), \; i\frac{\partial}{\partial t}\psi_{1} (x))] d^{3}{\bf x},
\label{45}
\end{equation}
is relativistically invariant.

Take into account the single valued decomposition of Cartan \cite {10}, 
\cite {11}
\begin{equation}
g = g_{p}g_{r},
\label{46}
\end{equation}
expressing any element $g$ of the group $L^{\uparrow}_{+}$ through an element  $g_{r}$ of the rotation groups $SO(3)$ and through the pure Lorentz transformation $g_{p}$ which converts, as well as the transformation $g$ does, the rest particle momentum $p_{0}$ into momentum $p$. (The pure Lorentz transformation leaves the axes of spatial coordinates of the new reference frame parallel to the axes of the initial reference frame, and thus, the real matrix $\Lambda$ from (\ref{4}) is Hermitian: $\Lambda^{\dagger}(g_{p}) = \Lambda(g_{p})$). Then any vector of  orbit (\ref{35}) can be written in the form
\begin{equation}
\psi_{gjk}(x) = \sum_{k'} c_{k'k} S(g_{p}) u_{jk'}(p_{0}) \exp(-ipx),
\label{47}
\end{equation}
where $c_{k'k}$ are numerical factors. Vectors $\psi (x)$ from the linear hull   ${\cal L}_{jp_{0}}$ represent a superposition of vectors (a wave packet) of orbit (\ref{47}), namely
\begin{equation}
\psi_{j}(x) = \sum_{k'} \int F_{k'}({\bf p}) S(g_{p}) u_{jk'}(p_{0}) 
\exp(-iEt+i{\bf p}{\bf x}) \frac{d^{3}{\bf p}}{2E},
\label{48}
\end{equation}
where $E = \sqrt{{\bf p}^{2}+M^{2}}$, $F_{k'}({\bf p})= \sum_{k} c_{k'k} 
f_{k}({\bf p})$, and $f_{k}({\bf p})$ are arbitrary square-integrable functions of three-momentum.

Substituting expression (\ref{48}) for $\psi_{2}$ and $\psi_{1}$ in formula (\ref{45}) and taking into consideration the relativistic invariance of the bilinear form in the abstract space ${\cal L}_{a}$ together with relations 
(\ref{17}) and (\ref{32}), we obtain sequentially
$$(\psi_{j}, \psi_{j})_{{\cal L}_{a} \otimes {\cal L}_{f}} 
= \sum_{kk'} \int F^{*}_{k}({\bf p})F^{}_{k'}({\bf p})
(S(g_{p}) u_{jk}(p_{0}),\; S(g_{p}) u_{jk'}(p_{0}))\frac{d^{3}{\bf p}}{2E}$$
\begin{equation}
=\left[ \sum_{\tau \in S(g)}w^{*}_{M\tau^{*}j} a_{\tau^{*}\tau}(j)w_{M\tau j}
\right] \sum_{k} \int |F_{k}({\bf p})|^{2} \frac{d^{3}{\bf p}}{2E}.
\label{49}
\end{equation}
It follows from here, that relativistically invariant bilinear form (\ref{49}) either has the same sign or is equal to zero for all vectors from the space 
${\cal L}_{jp_{0}}$. 

We take into account that any state of a field of class ISFIR or of class FSFIR, satisfying equations (\ref{27}) and (\ref{28}) and describing a particle with mass $M$ and rest spin $j$, possesses in the rest particle frame well-defined spatial parity $r$ which is equal to 1 or -1 (see, for example, 
\cite{19}). As the operator of the spatial reflection converts the 
$L^{\uparrow}_{+}$-group representation $\tau=(j_{0},j_{1})$ into the representation $\dot{\tau}=(-j_{0},j_{1})$, and as the finite-dimentional representations $\tau^{*}$ and $\dot{\tau}$ are equivalent, the equality 
\begin{equation}
w_{M\tau^{*}j} = r w_{M\tau j}
\label{50}
\end{equation}
is valid.

Relation (\ref{18}) gives for finite-dimentional representations of the group $L^{\uparrow}_{+}$ 
\begin{equation}
a_{\tau^{*}\tau}(j) = (-1)^{[j]}a_{\tau^{*}\tau},
\label{51}
\end{equation}
where the numeric values of $a_{\tau^{*} \tau}$ are arbitrary. In the framework of the given linear hull ${\cal L}_{jp_{0}}$ characterized by mass 
$M$, rest spin $j$ and parity $r$, we set the quantities $a_{\tau^{*} \tau}$ for all $\tau \in S(g)$ to the same value defined as
\begin{equation}
(-1)^{[j]}a_{\tau^{*}\tau}r = 1.
\label{52}
\end{equation}
For two different linear hulls the fixed values of the quantities 
$a_{\tau^{*} \tau}$ may be identical as well as different.

So, every Poincar\`{e} group representation realized in this or that linear space ${\cal L}_{jp_{0}}$ with definite values of mass and rest spin of a particle is unitary, because in such a space there exists a positive definite and Poincar\`{e}-invariant bilinear form. 

Advert now to giving arguments in favour of the irreducibility of the 
Poincar\`{e} group representation in the linear hull ${\cal L}_{jp_{0}}$. A not groundless doubt in respect to the irreducibility can be caused by that the representation $S(g)$ of the proper Lorentz group in the abstract space participating in the formation of the Poincar\`{e} group representation in the space ${\cal L}_{jp_{0}}$ is reduсible.

To clarify the matter of unusualness of the linear hull ${\cal L}_{jp_{0}}$, we first consider the situation with the linear hull  ${\cal L}_{u_{0}}$ generated by some vector of the $S(g)$-representation space
\begin{equation}
u_{0}  = \sum_{\tau \in S(g)} 
v_{\tau jk} \xi_{\tau jk}
\label{53}
\end{equation}
with fixed values of spin $j$, its projection $k$ and quantities $v_{\tau jk}$.
To confirm the reducibility of the representation in the linear hull 
${\cal L}_{u_{0}}$, it is enough to point out the existence of nonzero vectors in it with  zero values of the components associated with any given in beforehand irreducible representation $(j_{0}, j_{1}) \in S(g)$. In particular, these vectors are the ones specified by the formula  
\begin{equation}
u'_{0}  = \sum_{n=1}^{|j_{1}|-|j_{0}|} c_{n} \exp (iS^{30}\alpha_{n})u_{0},
\label{54}
\end{equation}
and the coefficients $c_{n}$ form a nontrivial solution of the equation system
\begin{equation}
\sum_{n=1}^{|j_{1}|-|j_{0}|} c_{n} A^{(j_{0},j_{1})}_{j'k,jk}(\alpha_{n})=0,
\label{55}
\end{equation}
where the boost parameters $\alpha_{n}$ are any real numbers, and $j'= |j_{0}|, |j_{0}|+1, \ldots |l_{1}|-1$.

In the superposition (\ref{33}) of vectors $\psi_{\alpha jk}(x)$ with various values of the boost parameter $\alpha$, belonging to linear hull 
${\cal L}_{jp_{0}}$, every summand has an extra multiplier 
$\exp (-ip_{\alpha}x)$ in comparison with the superposition (\ref{54}). With any choice of the superposition coefficients in the vector from 
${\cal L}_{jp_{0}}$, it does not allow to eliminate components associated with this or that irreducible representation $\tau \in S(g)$. The general assertion, that all vectors in the linear hull ${\cal L}_{jp_{0}}$ have components associated with every irreducible representation $\tau \in S(g)$, should be based on expression (\ref{48}).

Let's assume that in the space ${\cal L}_{jp_{0}}$ there exists a nonzero vector $\psi_{j} (x)$ (\ref{48}) containing no components associated with some representation $\tau_{0} \in S(g)$ giving nonzero contribution 
($w_{M\tau_{0}j} \neq 0$) to vector (\ref{32}). It means that there exists such a number $k_{0}$ and such a value of the three-momentum (the letter, by virtue of equivalence of all directions, can be considered as directed along the third axis and corresponding to the boost parameter value $\alpha$), that $F_{k_{0}}({\bf p}_{\alpha}) \neq 0$ and
\begin{equation}
(\xi_{\dot{\tau}_{0}j'k'}, \; \sum_{k}F_{k}({\bf p_{\alpha}}) 
\exp(iS^{30}\alpha) u_{jk}(p_{0}))  = 0
\label{56}
\end{equation}
for all values of spin $j'$ and its projections $k'$ contained in the representation $\tau_{0}$. Using relations (\ref{17}) and (\ref{33}), we obtain from (\ref{56})
\begin{equation}
F_{k'}({\bf p_{\alpha}}) A^{\tau_{0}}_{j'k',jk'}(\alpha) w_{M\tau_{0}j}  = 0.
\label{57}
\end{equation}
It follows from here that either $F_{k_{0}}({\bf p}_{\alpha})= 0$, or the condition (\ref{57}) is invalid, i.e., the aforestated assumption regarding the existence of a Poincar\`{e}-invariant subspace in the space 
${\cal L}_{jp_{0}}$ not coinsiding with ${\cal L}_{jp_{0}}$, is incorrect. It is a powerful enough argument in favour of the opinion on the irreducibility of the Poincar\`{e} group representation in the linear hull ${\cal L}_{jp_{0}}$.

Therefore, the linear space of normalized solutions of any of the considered relativistic-invariant equations (\ref{27}) with fields assigned to nonunitary representations $S(g)$ of the proper Lorentz group in the abstract space 
${\cal L}_{a}$ decomposes into a direct sum of spaces of unitary irreducible representations of the Poincar\`{e} group.

\begin{center}
{\bf 2.8. On Poincar\`{e} group representations not connected with the relativistic-invariant equations}
\end{center}

The fact that vector (\ref{32}) generating the linear hull ${\cal L}_{jp_{0}}$ is a solution of equation (\ref{27}), finds its reflection in three aspects: (1) the wave vector in the rest frame of a particle possesses a definite value of spin and, consequently, it is a proper vector of the Casimir operator $W$; (2) the wave vector in the rest frame of a particle possesses a definite value of the spatial parity, that provides the unitarity of Poincar\`{e} group representation in the linear hull ${\cal L}_{jp_{0}}$; (3) all the quantities $w_{M\tau j}$ in the wave vector (\ref{32}) have (up to a common factor) 
well-defined values given by each of the particular equations, that provides the irreducibility of Poincar\`{e} group representation in the linear hull ${\cal L}_{jp_{0}}$.

The above-stated approach to constructing the Poincar\`{e} group representations in the "hybrid" space ${\cal L}_{a} \otimes {\cal L}_{f}$ can be used without having attached to any linear relativistic-invariant equation and to any class of fields. For this purpose, it is necessary: first, to choose a $L^{\uparrow}_{+}$-group representation $S(g)$ in the space ${\cal L}_{a}$; second, to fix in this space some vector $u_{c} = \sum_{\tau\in S(g)} 
\sum_{jk} u_{\tau jk} \xi_{\tau jk}$; third, to fix at our discretion a 
four-vector of the momentum $p_{c}$; and, fourth, to form an orbit generated by acting with all possible Poincar\`{e} group transformations on the vector $u_{c}\exp (-ip_{c}x)$ of the space ${\cal L}_{a} \otimes {\cal L}_{f}$, and then to span the linear hull ${\cal L}_{u_{c}p_{c}}$ by this orbit. The set of Poincar\`{e} group representations in such linear hulls covers all conceivable types of states: massive $p_{c}^{2} > 0$, massless $p_{c}^{2} = 0$, tachyonic $p_{c}^{2} < 0$, and vacuum-like $p_{c} = 0$ ones. Evidently, one can formulate many mathematical problems concerning the representations in the spaces 
${\cal L}_{u_{c}p_{c}}$. However, we are not going to pay any attention to them, as it would mismatch the goals the present work.

\begin{center}
{\large \bf 3. Poincar\`{e} group representations of the Wigner's type}
\end{center}

\begin{center}
{\bf 3.1. On the key feature of the Wigner's realization of the Poincar\`{e} group representations}
\end{center}

As the time has shown, the most impressing aspect of Wigner's work \cite {8} was the inclusion of spin in some realization of the Poincar\`{e} group representations. It is notable that in \cite {8}, the functions of space-time coordinates $\varphi (x)$ are only considered in the context of translation transformations. Attention is focused on the proper vectors $\exp (-ipx)$ of the translation generators. The proper values of these generators, the four-vectors $p$, are declared to be the values of the momentum as of conservative quantity according to Noether's theorem. To construct  representations with nonzero spin, the Fourier transform of functions $\varphi (x)$ is introduced. The momentum $p$ as an argument of the Fourier image seems convenient in two respects: first, it does not change at space-time translations and, thereby, its transformations generated by the Poincar\`{e} group, are reduced to transformations generated by the proper Lorentz group; second, in realizing the Poincar\`{e} group representations, it is admissible to restrict oneself to a Lorentz-invariant domain $p^{2} = {\rm const}$ for the functions of the four-momentum $p$. (Analogous domains in the space-time, $x^{2} = {\rm const}$, are broken by translations). In the considered domain $p^{2} = M^{2} > 0$, there is one particular point $p_{0} = \{ M, 0, 0, 0 \}$ playing a crucial role in the Wigner's construction of the Lorentz group transformations of the spin.

In paper \cite{8} and then in monographs \cite{10}, \cite{11}, in the framework of constructing Poincar\`{e} group representations, the finite proper Lorentz group transformation corresponding to an element $g \in L^{\uparrow}_{+}$, is specified in the space of functions of the four-momentum $p$ and of a discrete variable $\varsigma$ as follows
\begin{equation}
\phi^{\prime}_{\varsigma} (p) = 
D^{j}_{\varsigma \varsigma'}(g_{w}) \phi_{\varsigma'} (\Lambda^{-1}(g)p)
\label{58}
\end{equation}
where $D^{j}(g_{w})$ is the matrix of transformation in the space of the rotation group representation with spin $j$ corresponding to the element 
$g_{w}$ assigned to the group $SO(3)$, which is expressed through the elements of the proper Lorentz group via the formula
\begin{equation}
g_{w} = g^{-1}_{p}gg^{}_{\Lambda^{-1}(g)p}.
\label{59}
\end{equation}
A detailed description of $D^{j}$-matrix elements as functions of three Euler's angles can be found in \cite{15}.

Relations (\ref{58}) and (\ref{59}) are the key relations in the concept of spin proposed by Wigner \cite{8} and in Wigner's construction of finite proper Lorentz group transformations leading to Poincar\`{e} group representation. Therefore, we shall pay peculiar attention to them.

The initial and final momenta in relation (\ref{58}), $\Lambda^{-1}(g)p$ and $p$, can be obtained from the rest momentum $p_{0}$ by pure Lorentz transformations corresponding, respectively, to the elements 
$g^{}_{\Lambda^{-1}(g)p}$ and $g_{p}$ of the group $L^{\uparrow}_{+}$. The transformation of $\Lambda^{-1}(g)p$ into $p$ generated by an element $g$ can be uniquely expressed through two consecutive operations: first, through the pure Lorentz transformation of momentum $\Lambda^{-1}(g)p$ into rest momentum  $p_{0}$, and then, through the transformation of momentum $p_{0}$ into $p$ corresponding to a uniquely found element $g_{1}$. 

In turn, according to Cartan decomposition of type (\ref{46}), the element $g_{1}$ can be expressed through two transformations. First of them is 
the rotation of coordinate axes $g_{w}$ which does not change the rest momentum $p_{0}$ and is adjusted in such a way that the second transformation converting
$p_{0}$ into $p$ be the pure Lorentz one. As a result, we have
\begin{equation}
g = g^{}_{p}g^{}_{w}g^{-1}_{\Lambda^{-1}(g)p}.
\label{60}
\end{equation}
Relation (\ref{59}) follows from here.

The above description of the element $g_{w}$ shows, first of all that, as the matter of fact, the transformation (\ref{58}) violates the equal status of all inertial reference frames, what is a fundamental essence of the relativity theory, the most significant property of Maxwell's electromagnetism theory, of quantum electrodynamics etc.

Consider some consequences of relations (\ref {58}) and (\ref {59}).
 
Let the proper Lorentz group element $g$ describes some rotation of the coordinate axes of any inertial reference frame without boosting it: $g = 
g_{r}$. One can show that
\begin{equation}
g^{}_{\Lambda^{-1}(g_{r})p} = g^{-1}_{r}g^{}_{p}g^{}_{r},
\label{61}
\end{equation} 
and, therefore, the element $g_{w}$ given by formula (\ref{59}) is identical to the geometrical rotation element $g_{r}$ at any value of the momentum $p$.
Due to that, relation (\ref{58}) reproduces the standard description of the representation of the rotation group $SO(3)$ in the tensor product of the spin space and the space of functions of the three-momentum ${\bf p}$.

Let $\Lambda^{-1}(g)p = p_{0}$, and an element $g$ correspond to the pure Lorentz transformation (the boost) $g = g_{p}$. Then the element  
$g_{\Lambda^{-1}(g)p}$ is the identity element $e$ of the group, and we obtain from (\ref{59}) that $g_{w}=e$ and $D^{j}_{\varsigma \varsigma'}(e)=
\delta_{\varsigma \varsigma'}$ in (\ref {58}). Consequently, in this case, the discrete index $\varsigma$, which can be identified with the spin projection onto the third axis, does not take part in the transformation of a wave vector. This situation contradicts the field transformation (\ref {33}), (\ref {34}) which are based on formulas (\ref{13}), (\ref{16}) obtained by Gelfand and Yaglom \cite{12}. In particular, it contradicts the transformation of the right $\psi_{R}$ and left $\psi_{L}$ components of Dirac fields, having for the boost along the third axis the following form
\begin{equation}
\psi^{\prime}_{R}(p) = \sqrt{\frac{E+|{\bf p}|}{M}}a_{+1/2}
\xi_{(1/2, 3/2)1/2\;+1/2}+\sqrt{\frac{E-|{\bf p}|}{M}}a_{-1/2}
\xi_{(1/2, 3/2)1/2\;-1/2},
\label{62}
\end{equation}   
\begin{equation}
\psi^{\prime}_{L}(p) = \sqrt{\frac{E-|{\bf p}|}{M}}b_{+1/2}
\xi_{(-1/2, 3/2)1/2\;+1/2}+\sqrt{\frac{E+|{\bf p}|}{M}}b_{-1/2}
\xi_{(-1/2, 3/2)1/2\;-1/2},
\label{63}
\end{equation} 
where $a_{k}$ and $b_{k}$ are constants. Relations (\ref{33}), (\ref{62}), 
(\ref{63}) demonstrate that every spin component of a wave vector boosted along the third axis changes by its individual factor. Also under the boost along any direction not parallel to the third axis, the mixing of the wave vector components with different spin projection values occurs, as it results from formulas (\ref{13})--(\ref {16}).

Let ${\bf p}$ be any three-momentum and an element $g$ correspond to a pure Lorenz transformation with its boost direction parallel to this momentum, i.e. to the boost related to the element $g_{p}$. Then the elements $g$ and $g_{p}$ commute with each other
\begin{equation}
gg_{p} = g_{p}g, \quad g_{p}g^{-1} = g^{-1}g_{p}.
\label{64}
\end{equation}
Write the momentum four-vector $\Lambda^{-1}(g)p$ in the form  
$\Lambda(g^{-1})\Lambda(g_{p})p_{0} = \Lambda(g^{-1}g_{p})p_{0}$. Using the hermicity of matrixes $\Lambda$ of the pure Lorentz transformations and relation (\ref{64}), we have
\begin{equation}
\Lambda^{\dagger}(g^{-1}g_{p}) = \Lambda^{\dagger}(g_{p})
\Lambda^{\dagger}(g^{-1}) = \Lambda(g_{p})\Lambda(g^{-1}) =
\Lambda(g_{p}g^{-1}) = \Lambda(g^{-1}g_{p}).
\label{65}
\end{equation}
It follows from here, that the element $g^{-1}g_{p}$ describes pure Lorentz transformation, and consequently, in the considered case $g_{\Lambda^{-1}(g)p}
= g^{-1}g_{p}$, and, as well as in the previous example, $g_{w}=e$.

It seems worthwhile to give now an example of any situation with the pure Lorentz transformation, when the right-hand side of formula (\ref{59}) gives 
a non-identity element $g_{w}$ refering to the group $SO(3)$. Identifying the elements of the group $L^{\uparrow}_{+}$ with 4$\times$4-matrixes $\Lambda$ and fixing the values of a four-vector $p$ and an element $g$, we find the values of all other quantities from relation (\ref{59}):
$$p_{0} = M \left( \begin{array}{c}
1 \\ 0 \\ 0 \\ 0 
\end{array} \right), \;
p = M \left( \begin{array}{c}
\sqrt{2} \\ 0 \\ 0 \\ 1 
\end{array} \right), \;
g = \left( \matrix {\sqrt{2} & 1 & 0 & 0 \cr 1 & \sqrt{2} & 0 & 0 
\cr 0 & 0 & 1 & 0 \cr 0 & 0 & 0 & 1} \right), \;
g_{p} = \left( \matrix {\sqrt{2} & 0 & 0 & 1 \cr 0 & 1 & 0 & 0 
\cr 0 & 0 & 1 & 0 \cr 1 & 0 & 0 & \sqrt{2}} \right),$$
$$\Lambda^{-1}(g)p = M \left( \begin{array}{c}
2 \\ -\sqrt{2} \\ 0 \\ 1 
\end{array} \right), \;
g_{\Lambda^{-1}(g)p} = \left( \matrix {2 & -\sqrt{2} & 0 & 1 
\cr -\sqrt{2} & 5/3 & 0 & -\sqrt{2}/3 
\cr 0 & 0 & 1 & 0 \cr 1 & -\sqrt{2}/3 & 0 & 4/3} \right),$$
\begin{equation}
g^{-1}_{p}gg^{}_{\Lambda^{-1}(g)p} =
\left( \matrix {1 & 0 & 0 & 0 \cr 0 & 2\sqrt{2}/3 & 0 & 1/3 
\cr 0 & 0 & 1 & 0 \cr 0 & -1/3 & 0 & 2\sqrt{2}/3} \right).
\label{66}
\end{equation}
We see, that the last matrix in (\ref{66}) is really equivalent to the matrix of rotation by an angle of $\arcsin (1/3)$ in the plane of the first and the third coordinate axes. Only due to the Lorentz transformations relating the present four-momenta $p$ and $\Lambda^{-1}(g)p$ with the fixed four-momentum, for which the rest momentum $p_{0}$ is chosen, it becomes possible (though not always, as the above examples show) to "convert" a pure Lorentz transformation $g$ into nontrivial rotation transformation $g_{w}$.

The matrix $D^{j}(g_{w})$ from relation (\ref{58}) becomes well-defined only after the element $g_{w}$ of the group $SO(3)$ is expressed in the standard form (see, for example, \cite{15}) through some analogues of Euler's angles.  Formula (\ref{58}) is applicable to any element $g$ of the proper Lorentz group, and so, as the above examples undoubtedly show, these analogues of Euler's angles depend on all the six, anyhow set, parameters of the group 
$L^{\uparrow}_{+}$ and, what is especially important, on the values of the argument (four-momentum) of the vector-valued function $\phi$ under transformation. In particular, infinitesimal values of parameters of an element $g_{w}$, defined by formula (\ref{59}) and corresponding to an element $g$ of infinitesimal transformation of the proper Lorentz group, are given by following relations
\begin{equation}
g = \left( \matrix {1 & \alpha_{1} & \alpha_{2} & \alpha_{3} \cr
\alpha_{1} & 1 & -\theta_{3} & \theta_{2} \cr
\alpha_{2} & \theta_{3} & 1 & -\theta_{1} \cr
\alpha_{3} & -\theta_{2} & \theta_{1} & 1} \right), \quad
g_{w} = \left( \matrix {1 & 0 & 0 & 0 \cr
0 & 1 & -\theta_{w3} & \theta_{w2} \cr
0 & \theta_{w3} & 1 & -\theta_{w1} \cr
0 & -\theta_{w2} & \theta_{w1} & 1} \right),
\label{67}
\end{equation}
with
\begin{equation}
\theta_{w1}=\theta_{1}+\frac{\alpha_{3}p^{2}-\alpha_{2}p^{3}}{E+M},  
\quad \theta_{w2}=\theta_{2}+\frac{\alpha_{1}p^{3}-\alpha_{3}p^{1}}{E+M}, 
\quad \theta_{w3}=\theta_{3}+\frac{\alpha_{2}p^{1}-\alpha_{1}p^{2}}{E+M}.
\label{68}
\end{equation}

As it is well-known, the definition of the group representation as a mapping
$g \rightarrow T(g)$ preserving the product operation contains no allusion to the admissibility or inadmissibility of any dependence of the parameters of the operator $T(g)$ on the characteristics of the vector of the representation space subjected to this operator. In our opinion, the identity of all parameters of the elements $g$ and of the operators $T(g)$ does naturally belong to the properties inherited by the representations from the group generating them. It is inherent, in particular, in the representations of the proper and orthochronous Lorentz groups found by Gelfand and Yaglom \cite{12}, in transformations of fields in relativistic-invariant equations (\ref{27}), and in the unitary irreducible representations of the Poincar\`{e} group obtained on the basis of the above two and referred to as the standard type. The violation of the mentioned identity, which is inherent in Wigner's realization \cite{8} of representations of the proper Lorentz group and of the Poincar\`{e} group, is accompanied by the violation of equal status of all elements of the proper Lorentz group in mapping them into the transformations fixed by relations (\ref{58}) and (\ref{59}). Apparently, it is of interest for the further mathematical constructions and generalizations, but leads to a number of losses in potential applications in physics.

\begin{center}
{\bf 3.2. On the nonexistence of the relativistic operator of spin in the 
Poincar\`{e} group representations of Wigner's type}
\end{center}

In contrast to the sentence about the dependence of parameters of the proper Lorentz group representation operator $T(g)$ given by formulas (\ref{58}) and (\ref{59}) on the value of the argument (four-momentum) of the function under transformation, a suggestion is likely to be made on the dependence of generators of the $L^{\uparrow}_{+}$-group representation on this argument, with keeping identity of all parameters of the elements $g$ and operators 
$T(g)$. It is possible to come to this suggestion if to split the three generators of the group $SO(3)$ in matrix $g_{w}$ (\ref {59}) in two families, making one of them connected with the parameters of rotations, and the other one with the parameters of boosts. As a result of such a procedure, the generators $\widetilde{L}^{\mu\nu}$ of the proper Lorentz group representation given by relations (\ref{58}) and (\ref{59}) can be expressed in the form found and discussed in works with participation of Foldy \cite{28}, \cite{29}:
\begin{equation}
\widetilde{L}^{\mu\nu}=\widetilde{M}^{\mu\nu}+\widetilde{S}^{\mu\nu},
\label{69}
\end{equation}
where
\begin{equation}
\widetilde{M}^{\alpha\beta}=i(p^{\alpha}\partial/\partial p_{\beta}
-p^{\beta}\partial/\partial p_{\alpha}), \quad 
\widetilde{M}^{\alpha 0}= iE\partial/\partial p_{\alpha},
\label{70}
\end{equation}
\begin{equation}
\widetilde{S}^{\alpha\beta} = S^{\alpha\beta}, \quad 
\widetilde{S}^{\alpha 0} = \frac{1}{E+M}\sum_{\gamma=1}^{3}p_{\gamma}
S^{\gamma\alpha},
\label{71}
\end{equation}
$E = \sqrt{{\bf p}^{2}+M^{2}}$, $\alpha, \beta = 1,2,3$.

The operators $\widetilde{M}^{\mu\nu}$ correspond to infinitesimal transformations of independent variables (the spatial components of a momentum) in the function $\phi$. They satisfy the standard commutation relations of type (\ref{9}) for generators of this or that proper Lorentz group representation. The antisymmetric tensor operator $\widetilde{S}^{\mu\nu}$ is responsible for infinitesimal transformations of the spin index of function $\phi$, whereas its pure spatial components coincide with the generators $S^{\alpha\beta}$ of the representation of the rotation group $SO(3)$ in the abstract (spin) space described by formulas (\ref{10})--(\ref{12}).

It is necessary to take into account that in the situation with nonzero spin the factor $(E+M)^{-1}$ in the parameters (\ref{68}) and in the generators 
(\ref{69})--(\ref{71}) leads to nonlocal character of infinitesimal Lorentz transformations of vector-valued functions of space-time coordinates 
$\varphi_{\varsigma}(x)$, with Furier images of the latter being the functions of momentum $\phi_{\varsigma}(p)$ subjected to transformation (\ref{58}).

It is easy to make sure that the components of antisymmetric tensor operator 
$\widetilde{M}^{\mu\nu}$ (\ref{70}) satisfy the commutation relations of type (\ref{9}) for $L^{\uparrow}_{+}$-group generators. At the same time, the six components of operator $\widetilde{S}^{\mu\nu}$ (\ref {71}) by themselves do not form a closed algebra. Besides that, they do not commute with the components of operator $\widetilde{M}^{\rho\sigma}$. In these conditions, it is of particular surprise that the components of the total operator 
$\widetilde{L}^{\mu\nu}$ (\ref{69}) reproduce the algebra of the proper Lorentz group (\ref{9}).

Therefore, the antisymmetric tensor operator $\widetilde{S}^{\mu\nu}$, as not being the generator of this or that $L^{\uparrow}_{+}$-group representation, cannot play the role of relativistic operator of spin as independent internal characteristic of a particle in any inertial reference frame. It has nontrivial consequences.

First, the classical relativistic description of spin expressed by the antisymmetric tensor $s^{\mu\nu}$ of Frenkel--Thomas \cite{3}, \cite{4}, has no quantum prototype in the Wigner's construction of the proper Lorentz group representations. Meanwhile, the description of spin by the tensor $s^{\mu\nu}$ entails the Thomas kinematic precession of spin \cite{4}, perfectly confirmed in precise experimental measurement of the muon $g-2$ \cite {30}. Together with the coupling of spin with magnetic moment proposed by Compton \cite{1} and Uhlenbeck and Goudsmit \cite {2}, this description has led to the equation found by Frenkel \cite{3} and then rediscovered by Bargmann, Michel, and Telegdi \cite{31} for the rotation of spin of a relativistic particle in a constant electromagnetic field, which is used, in particular, in polarization accelerator experiments (see, for example, \cite{32}).

Second, the operators $\widetilde{L}^{\mu\nu}$ cannot be separated into two independent families related to the relativistic orbital momentum and the relativistic spin, respectively, and so, it does not allow to reduce the 
Poincar\`{e} group representation of Wigner's type to the tensor product of representations of type (\ref{19}) and to separate the orbital (coordinate) and spin transformations of the wave vectors expressed by formula (\ref{8}). But the very relations (\ref{8}) and (\ref{19}) serve as the basis for formulating the conditions (\ref{23})--(\ref{26}), which the matrix operators from any linear equation (\ref{27}) should obey in order that this equation be relativistic invariant. These (and similar) conditions, supplemented with knowing the action of the $L^{\uparrow}_{+}$-group generators $S^{\mu\nu}$ on the basic vectors of this or that irreducible representation (\ref{10})--(\ref{16}), lead to exhaustive description (\ref{29})--(\ref{31}) of all linear relativistic-invariant equations (and lead in a straight way to various Lagrangians of field interactions).

Now it is worth noting a number of statements made in the work by Bargmann and Wigner \cite{9}. First, there is a repeatedly presented opinion borrowed from  
\cite{8} that "a classification of all unitary representations of the Lorentz group amounts to a classification of all possible relativistic wave equations". This opinion setting up false reference points in the field theory and particle physics research is completely refuted by the results of the work by Gelfand and Yaglom \cite{12} already commented in detail in section 2. Second, Bargmann and Wigner consider that in all cases the generators of the proper Lorentz group have the form $\bar{M}^{\mu\nu}+\bar{S}^{\mu\nu}$ where the operator $\bar{M}^{\mu\nu}$ acts on the variable $p$ and corresponds to the orbital angular momentum, while the operator $\bar{S}^{\mu\nu}$ acts on the variable $\varsigma$ and corresponds to the spin angular momentum, with both $\bar{M}^{\mu\nu}$ and $\bar{S}^{\mu\nu}$ satisfying the commutation relations of type (\ref{9}). Such a standpoint corresponds to the standard realization of irreducible representations of the proper Lorentz group which has been accomplished by Gelfand and Yaglom \cite{12}, and to the realization of Poincar\`{e} group representations of the standard type proposed above in section 2. But it has not found a confirming realization in \cite{9} as the operators $\bar{S}^{\mu\nu}$ are not concretized in any way, and, in view of the results of our discussion of formulas (\ref{58}), (\ref{59}), (\ref{69}) and (\ref{71}), it has prospects for its disproof because of its adherence to the Lorentz transformation
\begin{equation}
\phi^{\prime}_{\varsigma} (p) = 
Q_{\varsigma \varsigma'}(p, g) \phi_{\varsigma'} (\Lambda^{-1}(g)p),
\label{72}
\end{equation}
taken from \cite{8} and conceding dependence of the unitary operator $Q(p, g)$ on the momentum $p$.

In conclusion of this section we note that an important issue for particle physics, the question of the number of nonequivalent unitary irreducible representations of the Poincar\`{e} group of Wigner's type characterized by the same values of the Casimir operators $P$ (\ref{36}) and $W$ (\ref{37}), was not raised and was not discussed in \cite{8}--\cite{11}.

\begin{center}
{\large \bf 4. Towards the description of hadrons by reducible representations of the Poincar\`{e} group realized in space of ISFIR-class fields}
\end{center}

Both the parton bag model and the experimental picture of hadron states 
\cite {7} direct us to think that every stable hadron is accompanied by an infinite number of resonances. The quantum mechanical description of hydrogen atom teaches us that, along with the ground state of atom, the set of solutions of Schr\"{o}dinger equation inevitably contains all the excited states also. This lesson provides us with a serious argument in favour of the opinion, that
a justified correspondence between a stable hadron and a wave vector in any inertial reference frame given in beforehand is only possible in the case of  simultaneously reproducing all its resonances in the theoretical scheme.

We note first of all, that there is no direct way to establish a correspondance between those or other hadron states with known mass and rest spin values and suitable unitary irreducible representations of the Poincar\`{e} group of standard type, because, as it was noticed in section 2, the same mass and spin values of a particle refer to an infinite set of such representations. Besides that, in the space of vector-valued fields, we do not have any approach for the realization of reducible representations of the Poincar\`{e} group consisting of a given in beforehand set of irreducible representations, whatever it be. It is possible to provide finding various Poincar\`{e} group representations with the help of linear relativistic-invariant equations (\ref{27}), but the characteristics of irreducible representations composing them are unknown in advance, before the equations are solved.

As far as the simultaneous description of an infinite number of states with various mass and rest spin values is questioned, it is natural to study the diverse versions of the theory of infinite-component fields obeying equation (\ref{27}). Initially, such a study was concerned to FSIIR-class fields assigned to the $L^{\uparrow}_{+}$-group representations decomposable into a finite direct sum of infinite-dimensional irreducible representations. It has revealed certain features of the FSIIR-class field theory, listed in brief in section 2.4, which make it absolutely unsuitable for the particle physics.

If we start with the quark-gluon picture of hadrons, then, by decomposing the tensor product of Dirac spinors referring to valence and sea quarks, and 
four-vector fields referring to gluons, we obtain a direct sum of all finite-dimentional irreducible representations of the $L^{\uparrow}_{+}$-group with half-integer (for baryons) or integer (for mesons) spin. Nevertheless, the question of studying the theory of ISFIR-class fields which transform under the proper Lorentz group representations decomposable into an infinite direct sum of finite-dimensional irreducible representations, was not discussed in the literature till the 2000th. An essential obstacle for such studies, appart from mathematical complexity of arising problems, was an infinite number of arbitrary parameters in the equations for such class of fields given by relations (\ref{29})--(\ref{31}). 

Obtaining the efficient mechanism of selection of admissible 
$L^{\uparrow}_{+}$-group representations and eliminating the infinite arbitrariness in constants of the ISFIR-class field theory have been successfully performed in the framework of the double-symmetry notion whose strict formulation was given in \cite{33}. This notion includes, as particular cases, the $\sigma$-model symmetry of Gell-Mann and Levy \cite{34} and the supersymmetry. The starting building blocks of the double symmetry are the global group of the primary symmetry $G$ and some its representation $T$. The global or local group of the secondary symmetry has three marking properties: (1) it is generated by transformations whose parameters belong to the space of $G$-group representation $T$; (2) its transformations do not violate the primary symmetry; (3) it has no common elements with the group $G$, except for the identity element. The double symmetry is effecient as a method of constructing a field theory and can lead finally to a set of various groups of the secondary symmetry.

All versions of the free ISFIR-class field theory which along with the relativistic invariance (the primary symmetry) possesses also invariance under nontrivial global transformations of the secondary symmetry
\begin{equation}
\Psi '(x) = \exp [-i D^{\mu} \theta_{\mu}] \Psi (x),
\label{73}
\end{equation}
where parameters $\theta_{\mu}$ are components of a polar or axial four-vector of the orthochronous Lorentz group, and $D^{\mu}$ are matrix operators, have been found in \cite{17}. The closure of the algebra of operators 
$D^{\mu}$ is not questioned initially.

The existence of countable sets of versions of the double symmetric theory with polar four-vector parameters $\theta_{\mu}$ in transformations (\ref{73}) has been established. The fields of such a theory are assigned, in particular, to infinite-dimentional representations of the proper Lorentz group of the form
\begin{equation}
S^{s_{1}} = \sum^{+\infty}_{n_{1}=0} \sum^{s_{1}-1}_{n_{0}=-s_{1}+1}
\oplus (n_{0}, s_{1}+n_{1}),
\label{74}
\end{equation}
where the number $s_{1}$ are integer ($s_{1} \geq 1$) or half-integer ($s_{1} \geq 3/2$). For each value of the number $s_{1}$, the operator $R$ from equation (\ref{27}) is a multiple of the identity one: $R = \kappa E$; for the quantities $c(j'_{0},j'_{1};j_{0},j_{1})$ specifying the four-vector operator $\Gamma^{\mu}$ by means of relation (\ref{29}) the following equalities hold:
\begin{equation}
c(j_{0}+1,j_{1};j_{0},j_{1}) =c(j_{0},j_{1};j_{0}+1,j_{1})
=c_{0} \sqrt{\frac{(s_{1}-j_{0}-1)(s_{1}+j_{0})}
{(j_{1}-j_{0})(j_{1}-j_{0}-1) (j_{1}+j_{0}) (j_{1}+j_{0}+1)}},
\label{75}
\end{equation}
\begin{equation}
c(j_{0},j_{1}+1;j_{0},j_{1}) =c(j_{0},j_{1};j_{0},j_{1}+1)
=c_{0} \sqrt{\frac{(s_{1}-j_{1}-1)(s_{1}+j_{1})}
{(j_{1}-j_{0})(j_{1}-j_{0}+1) (j_{1}+j_{0}) (j_{1}+j_{0}+1)}},
\label{76}
\end{equation}
where $c_{0}$ is an arbitrary constant; the four-vector operator $D^{\mu}$ from the formula (\ref{73}) is a multiple of the operator $\Gamma^{\mu}$, $D^{\mu} = d_{0}\Gamma^{\mu}$, and due to that the secondary symmetry group is a four-parametrical abelian one, and the algebra of double symmetry group is isomorphic to Lie algebra of the Poincar\`{e} group.

The extension of the Lorentz group generated by the secondary-symmetry transformations (\ref{73}) entails, according to the Coleman--Mandula theorem \cite{35}, infinite spin degeneration of the mass spectrum of the free ISFIR-class field theory with the double symmetry. To avoid the degeneration one postulates spontaneous secondary-symmetry breaking at which the scalar (under the orthochronous Lorentz group) components of one or several bosonic ISFIR-class fields have nonzero vacuum expectation values $\lambda_{i}$, that can lead to changing the mass term of Lagrangian and the operator $R$ in equation (\ref{27}). To concretize such a change in equation (\ref{27}) for the fermionic fields, a strict solution to the problem of the existence and the structure of nontrivial fermion-boson interaction Lagrangians possessing the double symmetry is given in \cite{18}.

In a simple version of the fermionic ISFIR-class field theory with spontaneously broken double symmetry, when equation (\ref{27}) is described by the representation $S^{3/2}$ (\ref{74}) and by the constants (\ref{75}) and (\ref{76}) with $s_{1}=3/2$, the operator $R$ has the following form \cite{19}
\begin{equation}
R \xi_{(\pm\frac{1}{2},N+\frac{1}{2})jk} = \left[ \kappa + 
\sum_{i} 2\lambda_{i}q_{i}\frac{u_{i}^{N}(u_{i}N+N+1)-w_{i}^{N}(w_{i}N+N+1)}
{N(N+1)(u_{i}-w_{i})(2+u_{i}+w_{i})}\right] 
\xi_{(\pm\frac{1}{2},N+\frac{1}{2})jk}, 
\label{77}
\end{equation}
where $u_{i}=(z_{i}+\sqrt{z_{i}^{2}-4})/2$, 
$w_{i}=(z_{i}-\sqrt{z_{i}^{2}-4})/2$; $z_{i}$ are the free parameters describing the degree of the secondary symmetry breaking by a condensate of a given bosonic field (at $z_{i}=2$ the breaking is absent); $q_{i}$ are arbitrary constants. It is established \cite{19} that, in this version of the fermionic field theory, there is a broad range of free parameters $z_{i}$ and $\lambda_{i}q_{i}/c_{0}$ at which the mass spectrum has wonderful characteristics from the standpoint of hadron physics: (1) the mass spectrum is  bounded from below; (2) for each value of the rest spin and the spatial parity, there exists a countable set of mass levels extending up to infinity; (3) the lowest level mass value for a given rest spin increases as the spin increases; (4) the continuum part of the mass spectrum does not exist. It is also shown \cite{19} that in a situation with two parameters $z_{i}$, a satisfactory agreement between the theoretical levels and the experimentally observed nucleon resonances is attained.

The infinite number of levels in discussed versions of the free ISFIR-class field theory can be treated as a reflection of some internal structure of the corresponding particles, however, it is not possible to reexpress it in terms of the constituent elements and their interactions among themselves. It is remarkable that the considered theory does automatically assert the confinement, whereas, till now, it has the status of a hypothesis in the quantum chromodynamics. 

Now we can be assured enough that, in the framework of the ISFIR-class field theory with spontaneously broken double symmetry, it is possible to describe at an admissible level free hadron states of all sorts. For this purpose, it is necessary, first of all, to solve the general problem of variations in the mass terms in the bosonic-field Lagrangians which are caused by spontaneous breaking of the secondary symmetry and follow from three- and four-particle Lagrangians of self-interacting bosonic fields, taking into account the available variety of their internal quantum numbers.

The fundamental question which will arise in the further studies of the relativistic ISFIR-class field theory will concern the possibility of a satisfactory description of different types of hadron interactions by means of the monolocal Lagrangians and will demand solving a huge number of very interesting and extremely complicated problems, both mathematical and physical.

\begin{center}
{\bf Acknowledgments}
\end{center}

The author expresses sincere gratitude to S.P. Baranov, R.N. Faustov, I.P. Volobuyev, and B.L. Voronov, to ones for the discussions, induced me to prepare the present work, and to anothers for useful discussions on the considered problematics.

\end{small}
\end{document}